# HPHT growth of centimeter-sized cubic boron nitride crystals

Andrey Katrusha[a*], Weihua Peng[a], Jianguo Peng[a], Konstantin Iakoubovskii[b*]

[a] *Jilin Diamond Technology Research & Development, Jianan Road 177, Luyuan District, Changchun, China*
[b] *Central European Institute of Technology, Brno University of Technology, Purkyňova 123, 61200 Brno, Czech Republic*

**Abstract**
Single crystals of cubic boron nitride (cBN) exceeding 10 mm in size were grown by the high-pressure high-temperature (HPHT) temperature-gradient method using a Ni–Cr-based solvent catalyst. Compared with the previously reported maximum crystal size of approximately 3 mm, this improvement was achieved by maintaining a stable precursor flux during one week of growth at a source temperature of 1950 °C. In contrast to diamonds, which were grown with the same HPHT cell and showed nearly isometric shapes, the cBN crystals had elongated shapes. We attribute this cBN morphology to a localized growth near the BN source due to the relatively low effective mass transport of boron and nitrogen species in the metallic solvent.

*Corresponding authors: akatdps1@gmail.com, iakoubovskii@vut.cz

## 1. Introduction

Cubic boron nitride (cBN) is the closest rival of diamond in applications that require extreme hardness, high thermal conductivity and thermal stability, wide optical transparency down to deep ultraviolet range, and low dielectric constant. Such applications include windows for synchrotrons and high-power lasers, heat spreaders, and ultraviolet optoelectronic devices. Compared with diamond, cBN has slightly lower hardness and thermal conductivity and a somewhat higher refractive index, but it possesses a wider bandgap (see Table 1 [1-3]) and substantially higher oxidation resistance. Whereas bulk diamond begins to oxidize in air at approximately 800 °C, cBN remains stable above 1400 °C. [4].

Table 1. Comparison of basic properties of c-BN and diamond [1-4]: lattice constant *a*, refractive index *n*, band gap $E_g$, bulk modulus *K*, Knoop hardness, and thermal conductivity κ.

| **Material** | ***a* (nm)** | ***n*** | **$E_g$ (eV)** | **K (GPa)** | **Knoop (GPa)** | **κ (W/m·K)** |
|---|---|---|---|---|---|---|
| cBN | 0.3617 | 2.1 | 6.2 | 400 | 50-80 | 740 |
| Diamond | 0.3567 | 2.4 | 5.5 | 440 | 57-104 | 600-2000 |

Most of the applications mentioned above, especially those related to windows, require high-quality bulk single crystals, which can only be synthesized by the high-pressure high-temperature (HPHT) technique. Since the first report on the HPHT growth of cBN crystals with the maximum size reaching 0.3 mm in 1957 [5], only a ten-fold size increase has been achieved thus far [6], which is by far insufficient for applications. Crystal quality also remained limited, as indicated by relatively broad Raman lines, with the best reported values of 3-4 cm$^{-1}$ [7].

During the same period, HPHT diamond synthesis advanced to single crystals exceeding 28 mm in size [8], with Raman linewidths routinely reaching the intrinsic limit of approximately 1.5 cm$^{-1}$ in large crystals [9].

Several difficulties were mentioned in growing large cBN crystals. One was handling of the solvent catalyst, which typically contained an alkali or alkali earth metal and hence required handling in a dry inert atmosphere [10-12]. This issue was resolved by using an alkali-free Ni-based solvent, yet spontaneous secondary nucleation resulted in interaction among nearby crystallites and formation of irregular morphologies. This limited the synthesis time to ca. 3 hours at 1300–1700 °C in previous studies [11,12]. Finally, a higher temperature is required to convert hexagonal BN (hBN) into cBN as compared to the graphite-diamond transition (see Fig. 1). This difference originates from a higher activation energy for the conversion, and thus slower growth rates for cBN [5]. Due to the interaction among growing crystals and difficulties in maintaining relatively high temperatures and pressures, in the past experiments, either the duration was too short or the temperature was too low for growing crystals larger than 3 mm [11].

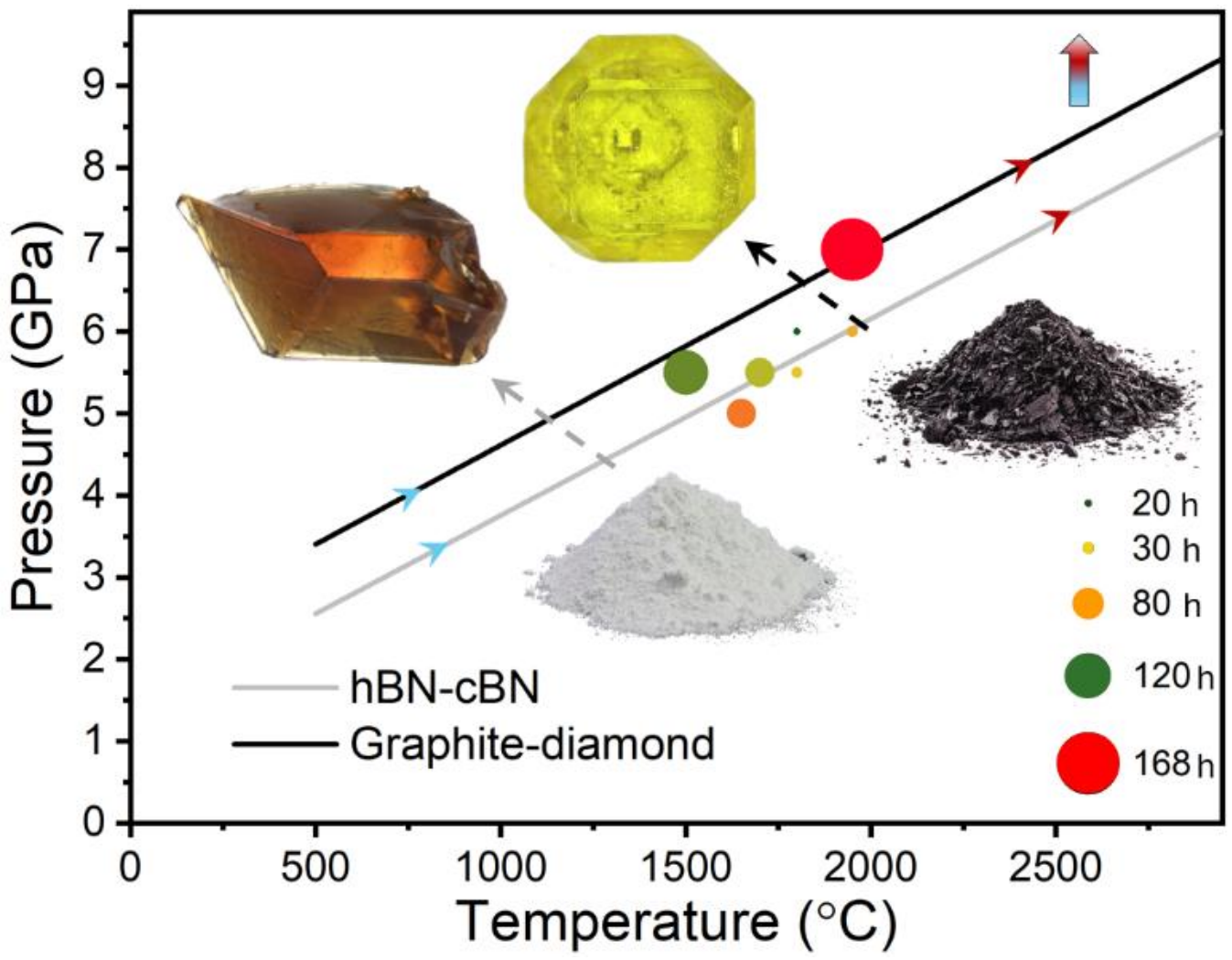

*Figure 1. Phase diagram for the hBN-cBN (gray line) and graphite-diamond conversion (black line); conversion rate increases with temperature as indicated by colored arrows [5]. Top photographs show representative morphologies for large diamond and cBN crystals: isometric for diamond and elongated for cBN. Symbols summarize the maximum pressure-temperature-duration parameters reported in the literature [7,10,13-15]. The red dot corresponds to this work.*

In this work, we optimized the HPHT cell geometry and solvent-catalyst composition and maintained stable growth for up to 168 h at 1950 °C. These improvements enabled the growth of cBN single crystals with a size exceeding 10 mm and a Raman linewidth as low as 1.8 cm$^{-1}$ (see Fig. 2)

## 2. Experimental details

Diamond and cBN crystals were synthesized at 6.5–7.0 GPa using the temperature gradient HPHT method and a cubic 6-anvil press designed in China. The growth duration was 48-168 hours. Diamond or cBN was used as the seed material. The seed region (see Fig. 3a) was heated up to 1600-1900 °C, while the source region was maintained at a higher temperature, producing a vertical temperature gradient of 1-5 °C/mm. Heat losses through the pressure medium and cell walls also produced a radial temperature gradient of approximately 3–12 °C/mm. Low gradient resulted in lack of growth (see center-left region in Fig. 3c), and hence the gradient distribution was carefully adjusted via the HPHT cell design to yield crystal nucleation on all seeds.

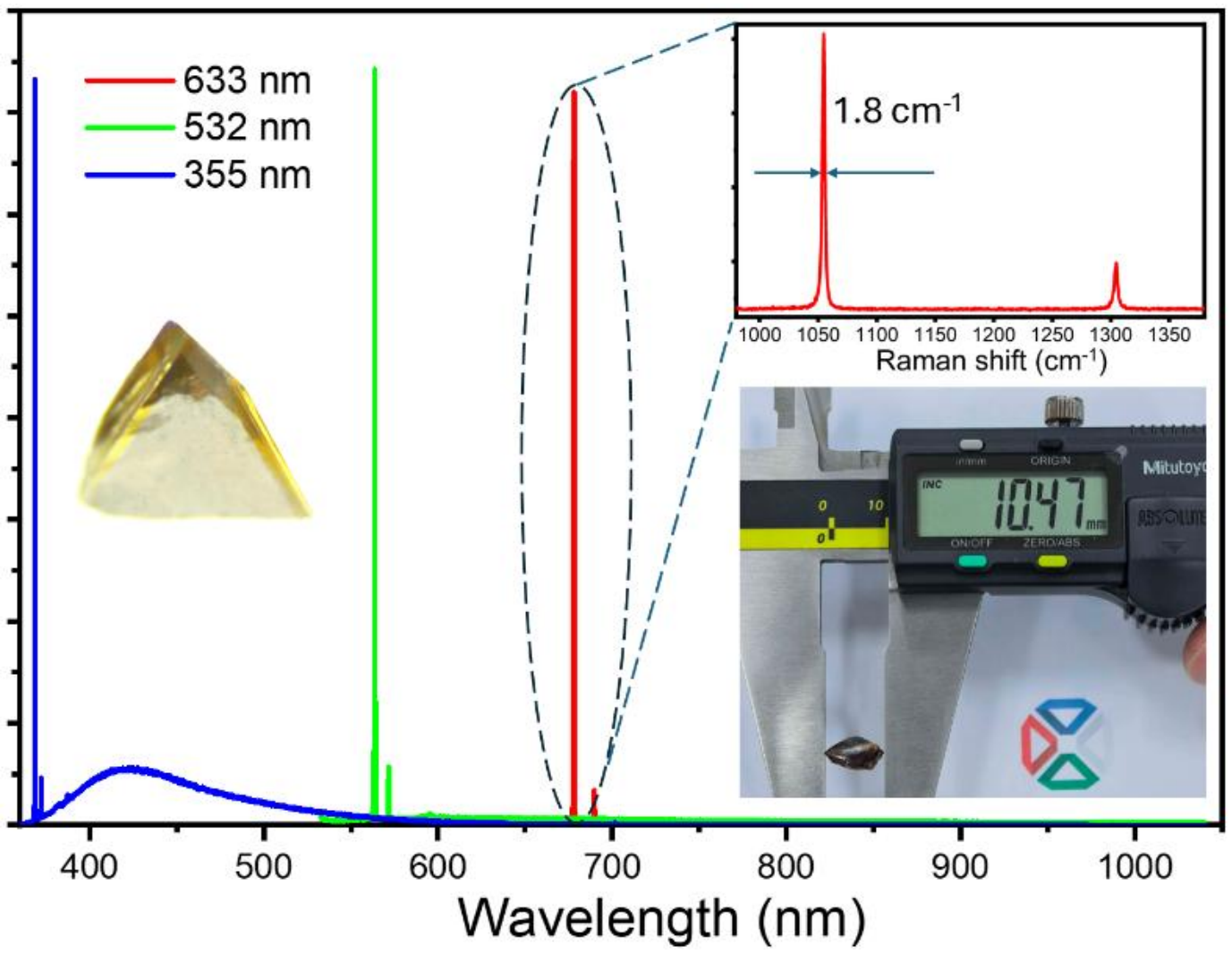

*Figure 2. Luminescence and Raman spectra of best cBN crystal synthesized in this work, measured under 633, 532 and 355 nm excitation. Inset shows a photograph of the largest crystal.*

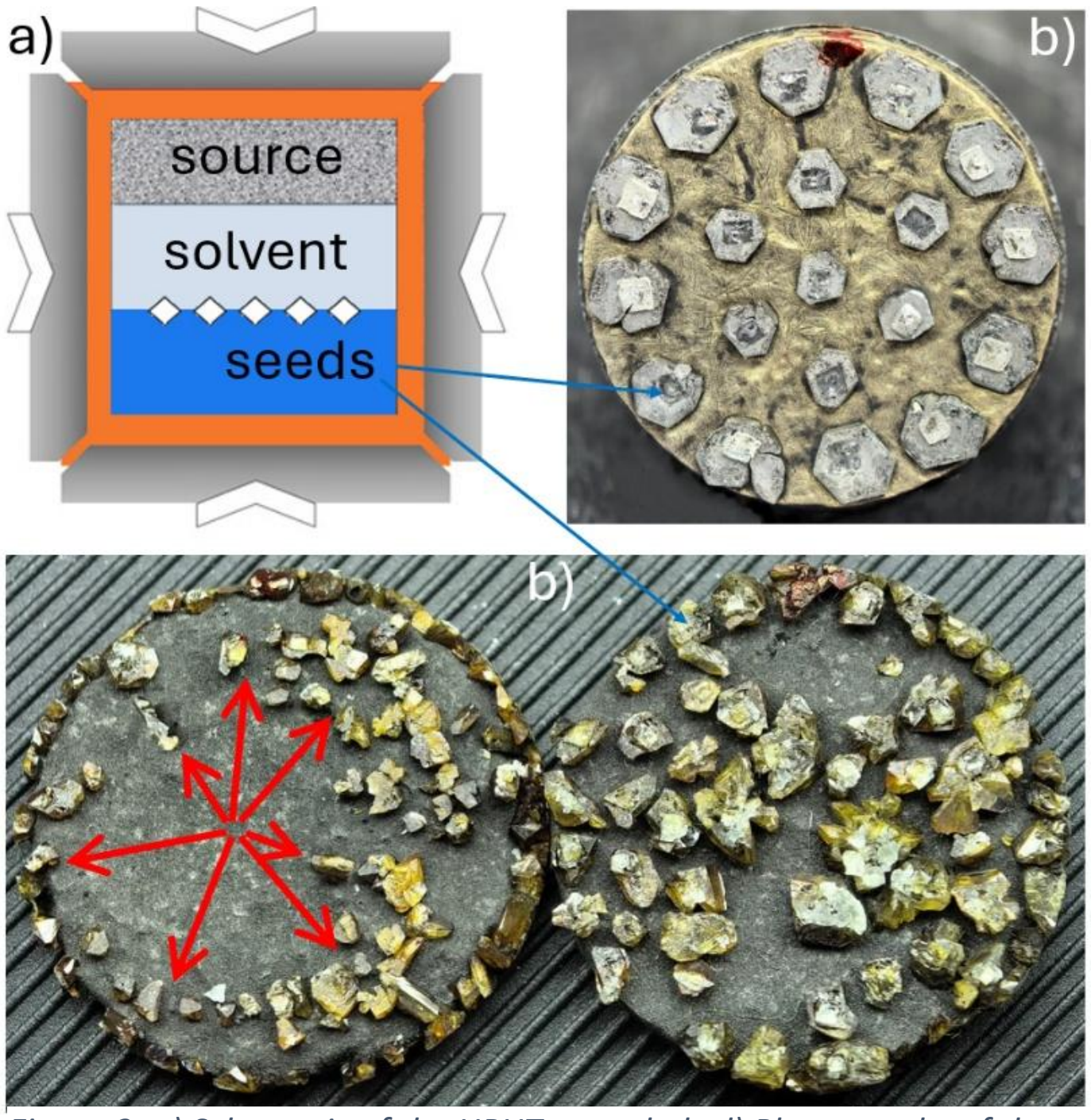

*Figure 3. a) Schematic of the HPHT growth. b-d) Photographs of the solvent with diamond (b) and cBN crystals (c-d). Small seed crystals can be discerned on some crystals. In c) the thermal gradient was low near the center resulting in lack of nucleation, while the shift of the symmetry axis from center to the left resulted from overheating of the left side of the cell that reduced the temperature gradient in the left part. Red arrows indicate the directions of lateral thermal gradients.*

High-purity (<0.30% oxygen) hexagonal boron nitride was used as the BN source. During the growth, it was dissolved in a Ni-based solvent-catalyst based on a Ni–Cr alloy (20-30 wt% Cr) with minor additions of Mg, Ti, Al, Si, or their combinations. The last four elements comprised less than 10 wt%; they acted as getters of oxygen, which is known to impart yellow color to cBN [16]. For diamond growth, we used a high-purity graphite source and a Fe-Co (65-35 wt%) catalyst. The source, solvent and growth zone were arranged in a geometry designed to maintain continuous liquid transport while reducing direct contact of the source with the region of crystal growth. The HPHT cell was primarily constructed from CsCl and phase-stabilized $ZrO_2$, selected for their mechanical and thermal stability under the applied conditions.

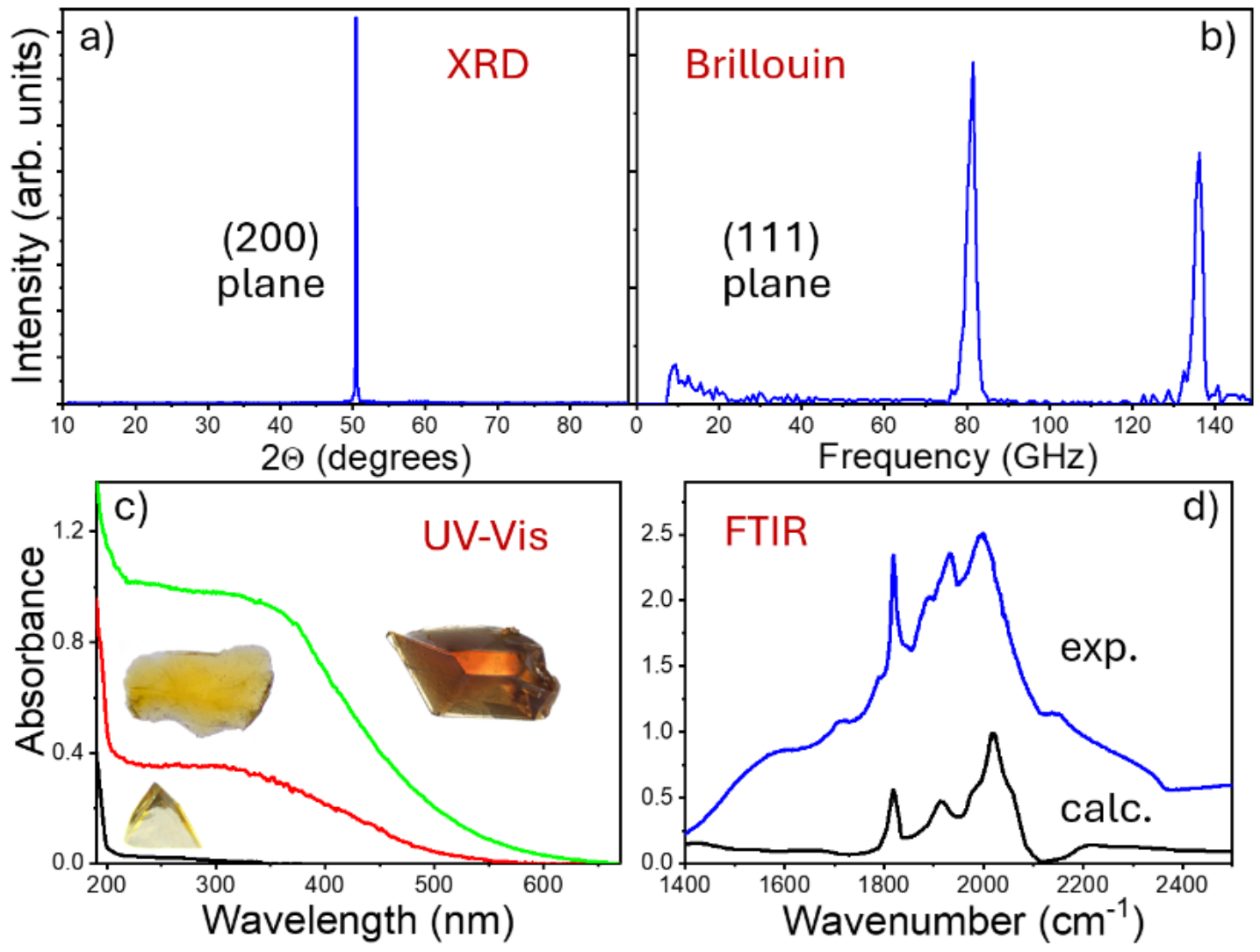

*Figure 4. Characteristic room-temperature spectra of reported cBN crystals: a) XRD, b) Brillouin, c) UV-visible and d) FTIR – experimental spectrum and calculated spectrum from ref. [17].*

The as-grown crystals were thoroughly cleaned using nitric acid, aqua regia and distilled water. Their phase composition was evaluated by Raman spectroscopy and X-ray diffraction. Their optical properties were reported elsewhere [16] and are partly shown in Figs. 2 and 4. Raman spectra were acquired with a WITec alpha 300R setup under 355, 532, or 632 nm excitation. For Brillouin spectroscopy we employed a TFP-2 spectrometer (JRS Scientific) equipped with a tandem 3+3 pass Fabry-Perot interferometer and a 532 nm laser. Fourier-transform infrared (FTIR) absorption spectra were recorded with a Vertex70v spectrometer from Bruker. Measurements were performed in vacuum to avoid atmospheric instabilities. For XRD we used a Rigaku MiniFlex 600 diffractometer and Ni-filtered CuKα radiation. All measurements were performed at ambient temperature.

## 3. Results and discussion

The inset photographs in Figure 1 show characteristic morphologies observed in this work: isometric for diamond and elongated for cBN. Both materials exhibited well-defined facets and sharp Raman peaks, with a full width at half maximum (FWHM) of 1.6 $cm^{-1}$ for diamond and 1.8 $cm^{-1}$ for cBN.

No additional phases were detected in cBN by Raman, XRD, Brillouin and FTIR spectroscopies as shown in Figs. 2 and 4: Raman spectra of Fig. 2 reveal peaks at 1055 and 1305 $cm^{-1}$ associated with the transverse optical (TO) and longitudinal optical (LO) modes of cBN. Luminescence is negligible under 355, 532 or 633 nm excitation, except for the dislocation-related [16] blue band excited at 355 nm.

The XRD pattern of Fig. 4a consists of a single peak at 50.45º associated with (200) crystal planes. Its FWHM = 0.1 º is rather narrow and is limited by the instrumental resolution.

The Brillouin spectrum of Fig. 4b shows two peaks at 81 and 136 GHz corresponding to transverse acoustic (TA) and longitudinal acoustic (LA) modes of a (111)-oriented cBN crystal. Within the experimental accuracies, their position matches those reported for high-quality cBN crystals grown at the National Institute of Materials Science in Japan. The resulting longitudinal sound velocity of 17.1 km/s is consistent with previous results and is only slightly below that of diamond (~17.5–18.6 km/s) [18].

UV-visible optical absorption spectra shown in Fig. 4c reveal a sharp onset near 200 nm, which corresponds to the bandgap absorption in cBN, together with oxygen-related broadband absorption that imparts most cBN crystals with yellow, amber or brown color [16]. Gettering of oxygen during the growth can yield colorless cBN crystals.

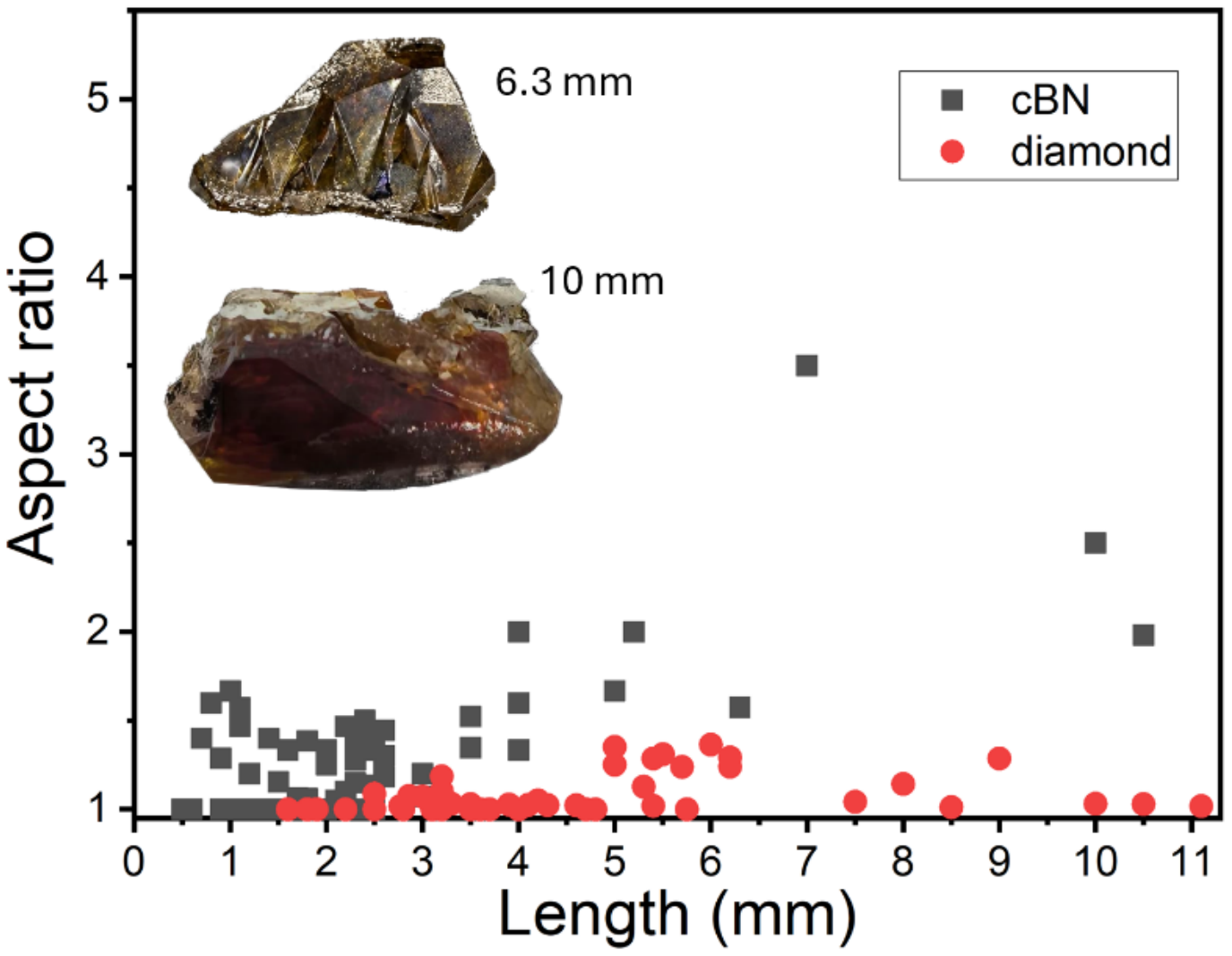

*Figure 5. Aspect ratio for cBN and diamond crystals grown in the same HPHT setup. Inset shows photographs of two cBN samples with indicated length.*

FTIR absorption (Fig. 4d) reveals an intrinsic two-phonon spectrum of cBN that matches rather well to the calculation results of reference [17]. No significant infrared-active impurities such as C–O, C=O, N-H or C-H could be detected.

To quantify the elongated shape of cBN crystals, in Figure 5 we plotted the aspect ratio, namely the ratio of the longest and shortest dimensions (length/width), as a function of crystal length. The aspect ratio quickly increases with crystal size for cBN, but not for diamond. Meanwhile, Figs. 3c demonstrates that the long axis of the cBN crystals is oriented approximately perpendicular to the vertical axis of the HPHT cell.

We tentatively explain this unusual cBN orientation as follows. At the high temperatures of HPHT synthesis, boron readily reacts with the components of the solvent, namely Ni, Cr, Ti and Al, while Ti and Al also readily react with nitrogen [19]. These reactions should limit effective mass transport of boron-, and possibly nitrogen-containing species in the solvent, resulting in BN accumulation near the source. Consequently, lateral flow of the source material produced by the unavoidable lateral temperature gradients may become comparable to or even exceed the vertical flow. The situation is different for diamond because of negligible reactivity of carbon with the Fe-Co solvent [19] and hence its faster vertical diffusion.

The observed elongation could in principle originate from either anisotropic crystal-growth kinetics or anisotropic mass transport. The fact that the morphology changes systematically with crystal size and that pronounced elongation occurs despite the use of a nominally symmetric cell suggests that transport limitations may play an important role; nevertheless, distinguishing transport from intrinsic facet-dependent growth kinetics requires further experiments.

## 4. Conclusions

We succeeded in growing high-quality single crystals of cubic boron nitride with a size exceeding 10 mm and Raman linewidth as narrow as 1.8 $cm^{-1}$. This was achieved by optimizing the HPHT cell geometry and the composition of the solvent-catalyst, avoiding the use of moisture-sensitive alkali-based alloys. The crystals exhibited pronounced elongation, which we tentatively attribute to limited effective mass transport of boron- and nitrogen-containing growth species through the metallic solvent. Future work will focus on developing solvent systems with lower chemical affinity for these elements in order to promote faster and more isotropic crystal growth.

## Acknowledgments

CzechNanoLab project LM2023051 funded by MEYS CR is gratefully acknowledged for the financial support of the measurements at CEITEC Nano Research Infrastructure. K.I. Is grateful to Jakub Krčma and Michal Urbánek for help with Brillouin measurements.